\documentclass[12pt]{article}
\usepackage{epsf}

\newlength{\dinwidth}
\newlength{\dinmargin}
\def\lapproxeq{\lower .7ex\hbox{$\;\stackrel{\textstyle <}{\sim}\;$}}
\def\gapproxeq{\lower .7ex\hbox{$\;\stackrel{\textstyle >}{\sim}\;$}}

\def\be{\begin{equation}}
\def\ee{\end{equation}}
\def\bea{\begin{eqnarray}}
\def\eea{\end{eqnarray}}

\def\GeV{{\rm GeV}}

\def\qq{{q\bar{q}}}
\def\gg{\gamma\gamma}

\def\slashq{/\!\!\!q}
\def\slashk{/\!\!\!k}
\def\slashe{/\!\!\!\epsilon}
\def\slashp{/\!\!\!p}

\begin{document}
\titlepage

\begin{flushright}
IPPP/04/55 \\
DCPT/04/110\\
3rd September 2004 \\
\end{flushright}

\vspace*{4cm}

\begin{center}
{\Large \bf Diffractive $\gg$ production at hadron colliders }

\vspace*{1cm} \textsc{V.A.~Khoze$^{a,b}$, A.D. Martin$^a$, M.G. Ryskin$^{a,b}$ and W.J. Stirling$^{a,c}$} \\

\vspace*{0.5cm} $^a$ Department of Physics and Institute for
Particle Physics Phenomenology, \\
University of Durham, DH1 3LE, UK \\[0.5ex]
$^b$ Petersburg Nuclear Physics Institute, Gatchina,
St.~Petersburg, 188300, Russia \\[0.5ex]
$^c$ Department of Mathematical Sciences, 
University of Durham, DH1 3LE, UK \\%
\end{center}

\vspace*{1cm}

\begin{abstract}
We compute the cross section for exclusive double-diffractive $\gg$ production
at the Tevatron, $p{\bar p} \to p+\gg+{\bar p}$, and the LHC.  We evaluate both the $gg$
and $\qq$ $t-$channel exchange contributions to the process.  The observation of
exclusive $\gg$ production at the Tevatron will provide a check on the
model predictions, and offer an opportunity to confirm the expectations for exclusive
double-diffractive Higgs production at the LHC.
\end{abstract}

\newpage

\section{Introduction}

\begin{figure}
\begin{center}
\centerline{\epsfxsize=0.5\textwidth\epsfbox{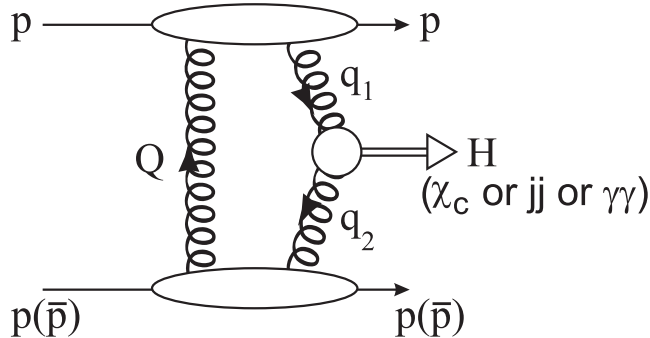}}
\caption{Schematic diagram for exclusive Higgs production at the LHC,
$pp \to p+H+p$.  The exclusive production of the states shown in the
brackets are possible `standard candles',
where the model predictions may, in principle, be checked by measurements
at the Tevatron.  Here we are concerned with the last process,
$p\bar {p} \to p+\gg+\bar {p}$.  Predictions already exist for the
exclusive production of Higgs bosons \cite{KMR, KMRProsp, KKMRext},
 $\chi_c$,$\chi_b$ \cite{KMRmm, KMRS}
 and dijets \cite{KMRdijet, KMR, KMRProsp}. }
\label{fig:H}
\end{center}
\end{figure}

The experimental study of the central exclusive double-diffractive production
processes at the Tevatron is interesting
in its own right, since it is an ideal way to improve our
understanding of diffractive processes and the
dynamics of the Pomeron exchange.   Moreover such observations can provide a
valuable check of the theoretical
models and experimental methods which may be used to search for the
new physics at LHC \cite{KMRProsp}. Of particular interest is
exclusive Higgs boson production
$pp\to p+H+p$ \cite{DKMOR}.
The + signs are used to denote the presence of large rapidity gaps; here
we will simply describe such processes as `exclusive', with
`double-diffractive' production being implied.   The predictions for
exclusive production are obtained by calculating the diagram of Fig.~\ref{fig:H}
using perturbative QCD.  In addition we have to calculate the probability
that the rapidity gaps are not populated by secondaries from the underlying
event.

However it is not easy to find an exclusive process which may be observed at the Tevatron
and so act as a `standard candle' for the theoretical predictions for exclusive
Higgs production.  Let us consider the possibilities.  These are shown in
brackets in Fig.~\ref{fig:H}.

Recently the first `preliminary' result on exclusive $\chi_c$ production has been
reported \cite{CDFchi}. Although it is consistent with perturbative QCD expectations \cite{KMRmm,KMRS},
the mass of $\chi_c$-boson, which drives the scale of the process, is too
low to justify just the use of perturbative QCD\footnote{Even lower scales
correspond to the fixed-target central double diffractive meson resonance production
observed by the WA102 collaboration at CERN \cite{WA102}.  Therefore,
it is intriguing that the qualitative features of the observed $p_t$ and
azimuthal angular distributions appear to be in good agreement with the
perturbatively based expectations \cite{KMRtag}.}.

One possible process with a larger scale is the exclusive production
of a pair of high $E_T$ jets, $p\bar {p} \to p+jj+\bar {p}$ \cite{KMRdijet, KMR, KMRProsp}.
In principle,
this  process appears to be an ideal `standard candle'.  The expected cross section is rather large,
and we can study its behaviour as a function of the mass of the dijet
system.  Unfortunately in the present CDF environment, the
background from the `inelastic Pomeron-Pomeron collisions' contribution is large as well.
Theoretically the exclusive dijets should be observed as a narrow peak,
sitting well above the background, in the
distribution of the ratio
\be
R_{jj}=E_{{\rm dijet}}/E_{\rm {PP}}
\ee
 at $R_{jj}=1$, where $E_{\rm {PP}}$ is the energy of the incoming
 Pomeron-Pomeron system.  In practice
the peak is smeared out due to hadronization and the jet-searching algorithm.
For jets with $E_T=10$ GeV and a jet cone $R<0.7$, more than 1 GeV will be lost
outside the cone, leading to (i) a decrease of the measured jet energy of about 1-2 GeV\footnote{
Note that the jet $E_T$ were not corrected in the preliminary data presented
in Ref.~\cite{CDFchi, CDFdijet}.},
and, (ii) a rather wide peak ($\Delta R_{jj}\sim \pm 0.1$) in the $R_{jj}$ distribution.
The estimates based on Ref.~\cite{KMRProsp} give an exclusive cross section for dijet
production with $E_T>25$ GeV (and CDF cuts) of about 40 pb, which is very close
to the recent CDF measurement \cite{CDFchi, CDFdijet},
\be
\sigma(R_{jj}>0.8,~E_T>25 ~{\rm GeV})~~=~~34~\pm 5({\rm stat}) ~\pm 10({\rm syst}) ~{\rm pb}.
\ee
Bearing in mind the large uncertainties (in both the theoretical calculations and in experimental
identification of low $E_T$ jets) at low
scales, the predictions \cite{Liverpool, KMRProsp} for $E_T > 7, 10$ GeV
are also in agreement
with the corresponding CDF measurements \cite{CDFjj,CDFchi, CDFdijet}.
 However there is no `visible' peak in the CDF data for $R_{jj}$ close to 1. The contribution
 from other channels (called `central inelastic' in Ref.~\cite{KMRProsp}) is too large, and
  matches with the expected peak smoothly\footnote{We hope that applying the $k_t$
  jet searching algorithm, rather than the jet cone algorithm, would improve the
 selection of the exclusive events.   This is in accord with the studies in Ref.~\cite{CFP}.}.

\begin{figure}
\begin{center}
\centerline{\epsfxsize=0.8\textwidth\epsfbox{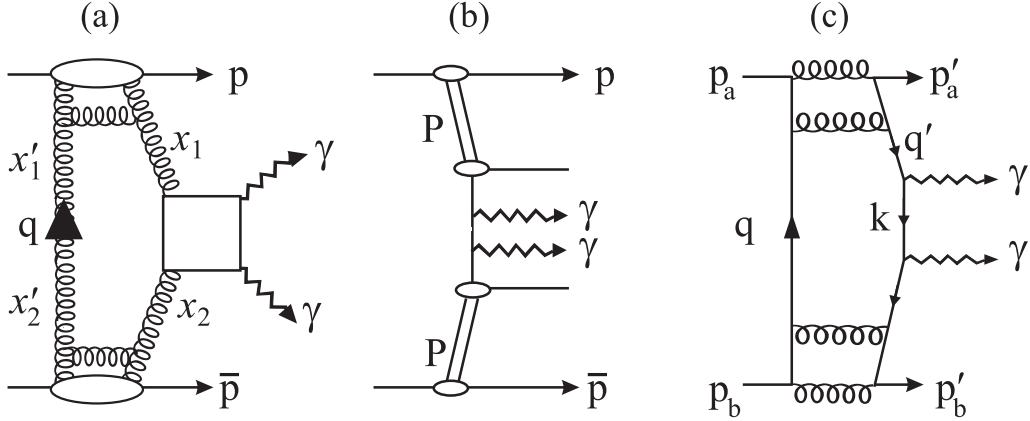}}
\caption{(a) Exclusive $\gg$ production driven by two-gluon
$t$-channel exchange, and the backgrounds arising from, (b)
$\gg$ produced in a Pomeron-Pomeron initiated subprocess,
accompanied by soft undetected hadrons
and, (c) from $\gg$ production via $t$-channel quark-antiquark
exchange.  All permutations of the particle lines are implied.}
\label{fig:gg}
\end{center}
\end{figure}

An alternative possibility is to measure exclusive $\gamma\gamma$ production
with high $E_T$ photons, $p\bar {p} \to p+\gg+\bar {p}$ \cite{AR, KMRProsp}.
Here there are no problems with hadronization or with the identification of the jets.
Moreover, we can access much higher masses of the centrally
produced system than in the $\chi_c$ case.
On the other hand the exclusive cross section is rather small. As usual, the
perturbative QCD Pomeron is described by two (Reggeized) gluon exchange.
 However the photons cannot be emitted from the gluon lines
directly. We need first to create quarks. Thus a quark loop is required
(see Fig.~\ref{fig:gg}a), which causes an extra coupling
 $\alpha_s(E_T)$ in the amplitude.
The prediction of the cross section for
 exclusive $\gg$ production, and the possible background contributions,
 are the subject of this paper.

In order to isolate the component of exclusive $\gg$ production
which is driven by two-gluon $t$-channel exchange, we need to consider
other possible sources of these events.  Possibilities are:\\

 (i)  inclusive
 reactions in which the production of a $q\bar q$ pair is such that
  the quarks transfer almost the whole of their energy
 to the emitted photons (Fig.~\ref{fig:gg}b),
 so that any additional hadrons (coming from the hadronization of the $q$
 and $\bar q$) are
 soft, and so may be missed by the Central Detector;\\

(ii) diagrams with the $t$-channel quark exchange (Fig.~\ref{fig:gg}c).
We would expect this contribution to be suppressed at high energies.
The quark densities generated from the incoming valence quarks in a
fixed-order graph like Fig.~\ref{fig:gg}c behave as $x_i q(x_i) \sim x_i$,
whereas the gluons generated by the fixed-order diagram of Fig.~\ref{fig:gg}a have distributions
that behave as $x_i g(x_i) \sim$ constant,
modulo log$(x_i)$ factors.
However the parton distributions at low $x$ and moderate scales indicate
that the quark densities are comparable to that of the gluon.
On the other hand, the photons can be emitted directly from a quark line
 (without an extra loop and its accompanying small $\alpha_s$ factor).
 Moreover the `skewed' factor, $R_q$, due to the asymmetric $q\bar q$
 $t$-channel configurations is
 much larger for quarks, $R_q\sim 3-4.5$, than the corresponding factor for skewed $t$-channel gluons \cite{SGMR}.
 Since the exclusive cross section is proportional to $R_q^4~$ \cite{KMR}, this is clearly
 important.\\
\vspace {1cm}

\section{Exclusive $\gg$ production via $gg$ $t$-channel exchange}

First, we calculate exclusive $\gg$ production arising from gluon-exchange,
as shown in Fig.~\ref{fig:gg}a.
We write the cross section in the factorized form \cite{KMRProsp}
\begin{equation}
\label{eq:a2}
 \sigma_g \; = \; {\cal L}_g (M_{\gg}^{2}, y) \: \hat{\sigma}_g (M_{\gg}^{2}),
\end{equation}
where $\hat{\sigma}_g$ is the cross section for the hard $gg \to \gg$ subprocess
which produces a $\gg$ system of mass $M_{\gg}$, and ${\cal L}_g$ is the
effective $gg$ luminosity for production of a central system ($\gg$ in our case)
with rapidity $y$.
For the exclusive $\gg$ production shown in Fig.~\ref{fig:gg}a we have, to single
$\log$ accuracy, \cite{KMR}
\begin{equation}
\label{eq:a4}
  M_{\gg}^2 \: \frac{\partial {\cal L}_g}
 {\partial y \:
\partial M_{\gg}^2} \; = \; \hat{S_g}^2  \; \left ( \frac{\pi}{(N_C^2 - 1) b} \int \frac{dq_t^2}{q_t^4}
 \: f_g (x_1, x_1^\prime, q_t^2, \mu^2) \: f_g (x_2, x_2^\prime,
 q_t^2, \mu^2) \right )^2,
\end{equation}
where $b$ is the $t$-slope corresponding to the momentum transfer
distributions of the colliding proton and antiproton
\begin{equation}
\label{eq:a5}
 \frac{d^2 \sigma}{dt_1 dt_2} \; \propto \; e^{b (t_1 + t_2)}.
\end{equation}
We take $b =
4~{\rm GeV}^{-2}$. The quantities $f_g (x, x^\prime, q_t^2,
\mu^2)$ are the generalised (skewed) unintegrated gluon densities.
The skewed effect arises because the screening
gluon ($q$) carries a much smaller momentum fraction $x^\prime
\ll x$. For small $|x - x^\prime|$ the skewed unintegrated density
can be calculated from the conventional integrated gluon $g (x,
q_t^2)$ \cite{MR}. However the full prescription is rather
complicated. For this reason it is often convenient to use the
simplified form \cite{KMR}
\begin{equation}
\label{eq:a6}
 f_g (x, x^\prime, q_t^2, \mu^2) \; = \; R_g \: \frac{\partial}{\partial \ln
 q_t^2}\left [ \sqrt{T_g (q_t, \mu)} \: xg (x, q_t^2) \right ],
\end{equation}
which holds to 10--20\% 
accuracy.\footnote{In the actual computations
we use a more precise form as given by Eq.~(26) of \cite{MR}.}
The factor $R_g$ accounts for
the single $\log q^2$ skewed effect \cite{SGMR}.  It is found to
be about 1.4 at the Tevatron energy.
The Sudakov factor $T_g (q_t, \mu)$ \cite{Kimber,WMR} is the survival probability that
a gluon with transverse momentum $q_t$ does not emit any partons in the
evolution up to the hard scale $\mu = M_{\gg}/2$
\begin{equation}
\label{eq:a7}
 T_g (q_t, \mu) \; = \; \exp \left ( - \int_{q_t^2}^{\mu^2} \:
 \frac{\alpha_S (k_t^2)}{2 \pi} \: \frac{dk_t^2}{k_t^2} \:
 \int_0^1 \: \left [\Theta(1-z-\Delta)\Theta(z-\Delta)zP_{gg} (z) \: + \: \sum_q \:
 P_{qg} (z) \right ] \: dz \right ),
\end{equation}
with $\Delta = k_t/(\mu + k_t)$.  The square root arises in
(\ref{eq:a6}) because the survival probability is only relevant to
the hard gluon.  It is the presence of this Sudakov factor which
makes the integration in (\ref{eq:a4}) infrared stable, and
perturbative QCD applicable.   We use the MRST99 partons \cite{MRST99}, and cut the
loop integral at $q_t\geq 0.85$ GeV,
 as in \cite{KMRS}.

We also have to compute the probability, $\hat{S_g}^2$,  that the rapidity gaps
are not populated by secondaries from soft rescattering from the colliding
proton and antiproton.  We calculate $\hat{S_g}^2$ using a two-channel
eikonal model \cite{KMRsoft}.

\begin{figure}
\begin{center}
\centerline{\epsfxsize=0.8\textwidth\epsfbox{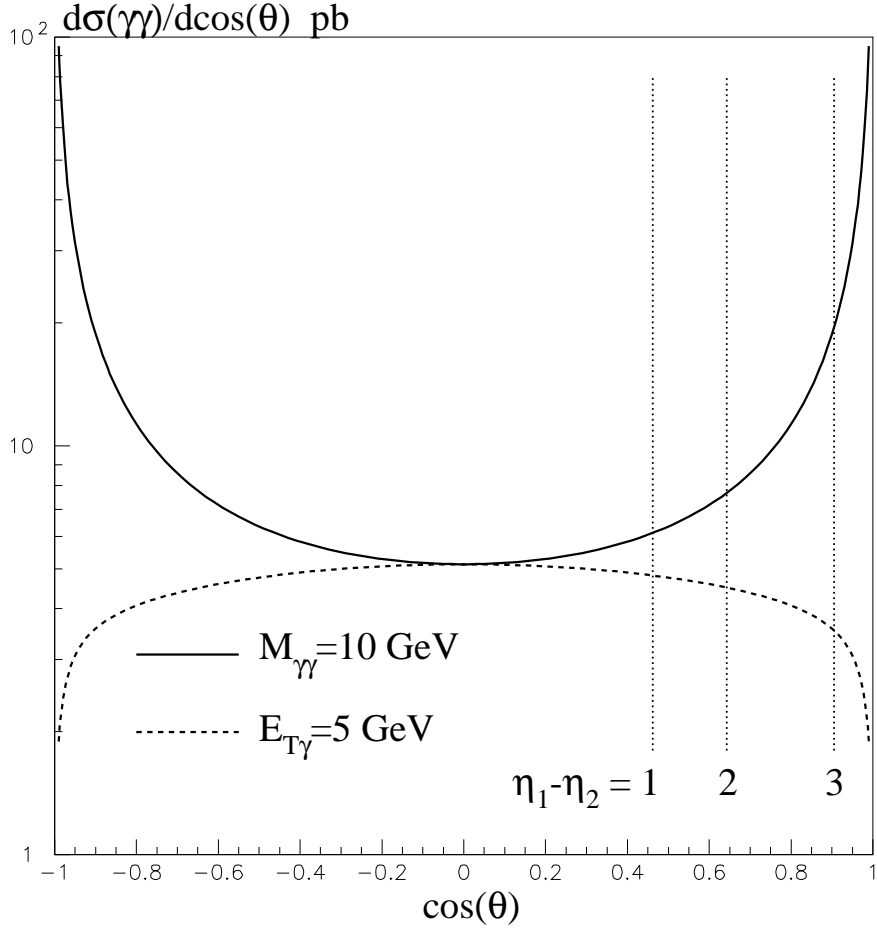}}
\caption{The behaviour of the differential cross section for the $J_z=0$ hard
subprocess $gg \to \gg$ in exclusive $\gg$ production, shown by the continuous
curve for fixed $M_{\gg}=10$ GeV
and by the dashed curve for fixed $E_{T\gamma}=5$ GeV.  The values of ${\rm cos}\theta$
corresponding to the
rapidity differences of the two emitted photons $\eta_1-\eta_2=1,2$ or 3
are indicated.}
\label{fig:sig}
\end{center}
\end{figure}

To compute the subprocess cross section, $\hat{\sigma}$,we use the known
QED results for the
 $\gamma\gamma\to\gamma\gamma$ helicity amplitudes from Refs.~\cite{[37], BFDCW}.
Note that the incoming active gluons are 
in a $P$-even, $J_z=0$ state \cite{Liverpool, KMRmm, KMRProsp}, where $z$ is
the proton beam direction.   Thus we need to compute the $J_z=0$ $~gg \to \gg$ cross
section, rather than the
usual cross section averaged over the gluon polarisations.
The results are shown in Fig.~\ref{fig:sig} for {\it fixed} $M_{\gg}=10$ GeV
(continuous curve) and for {\it fixed} $E_{T\gamma}=5$ GeV (dashed curve).
All four flavours of quark $(u,d,s,c)$ in the fermion loop were taken to
be massless. The vertical dotted lines indicate the angles corresponding
to a rapidity difference between the two photons $\eta_1-\eta_2=1,2$ or 3.
We see that the logarithmic enhancement of the differential cross section at
$|{\rm cos}~\theta| \to 1$ for fixed $M_{\gg}$ is strongly suppressed, by
the Jacobian, when we select events with fixed $E_{T\gamma}$.  The cross
section decreases, since the fixed value of $E_{T\gamma}$ can only be achieved at small angles by
increasing the value of $M_{\gg}$, while the cross section behaves as
$d\sigma/d{\rm cos}\theta \sim 1/M_{\gg}^2$.
Finally, combining the effective luminosity with the subprocess cross section,
we obtain the predictions for the exclusive $\gg$ cross section which we present, and discuss,
in Section 4.

There may be some contribution from the semi-elastic
reaction with forward proton dissociation. Such a contribution was discussed
 in \cite{KMRS}  for exclusive $\chi$ production. Provided the mass
 of the centrally produced system is not too large, it was argued that this
 contribution is small.\footnote{This reflects the smallness of the triple-Pomeron vertex
(see, for example, \cite{ABKK})  in soft
 processes. For the Tevatron the yields of exclusive and inclusive events are expected to be comparable
 at $M \sim 15$ GeV, when gaps $|\Delta\eta_{\rm gap}|>3$ are imposed in the inclusive case.}
Moreover, the CDF measurement selects events without any secondaries in the
pseudorapidity interval $3.5 < |\eta| <7.5$. This selection strongly
suppresses the possibility of forward proton dissociation, and the admixture of processes
with incoming proton dissociation is not expected to exceed 0.1\%.

 The background from the inclusive $q\bar q$ plus $\gamma\gamma$ production process, shown in Fig.~\ref{fig:gg}b,
may be estimated using the POMWIG Monte Carlo programme \cite{POMWIG,CFH}. It is not anticipated to be large.
For $E_T>12$ GeV photons, the whole cross section in the rapidity interval $|\eta_\gamma|<2$ ($\sim$ 100 fb) exceeds
 the exclusive cross section (Fig.~\ref{fig:gg}a) by a factor of about 50. However the probability
not to observe any hadrons produced via $q\bar q$
hadronization is very small, so we hope that this background can be suppressed sufficiently to see
the exclusive signal.

\section{Exclusive $\gg$ production via $\qq$ $t$-channel exchange}

Here we discuss the $\qq$ $t$-channel exchange contribution to
exclusive $\gg$ production, as shown in Fig.~\ref{fig:gg}c.
This contribution has some novel features, and so we discuss
its computation in detail.

At first sight it appears that the contributions of the $\qq$ exchange
graphs, Fig.~\ref{fig:gg}c, may be neglected immediately, since the amplitudes
are suppressed by the power factor $1/s$
in comparison with the $gg$ exchange graphs.  To be more precise, the suppression is
given by $x \sim$ exp$(-\Delta \eta_{\rm gap}$).
However, as mentioned in the Introduction, we must take care.    First, the amplitude for
the main process, Fig.~$\ref{fig:gg}$a, contains a factor of $\alpha _s(E_T)$ arising from the
quark loop.  Second the $\qq$-exchange contribution to the cross section
is enhanced by the skewed effect, $R_q^4 \sim 200$ \cite{SGMR}.   Third, at the relatively low
scales (few $~\GeV ^2)$, relevant for the exclusive production of a system of
mass $M \sim 10-30~\GeV$ at the Tevatron, the global (CTEQ \cite{CTEQ}, MRST \cite{MRST}) parton analyses
find valence-like gluons ($xg$ decreases as $x \to 0$, contrary to naive perturbative QCD
expectations) but Pomeron-like unpolarised singlet
quarks ($xq \sim x^{-\lambda}$ with $\lambda >0$).  For these reasons the
$\qq$-exchange contribution, Fig.~$\ref{fig:gg}$c, must be evaluated.

In analogy with the computation of the gluon-exchange contribution, (\ref{eq:a2}),
we write the quark contribution as the product of the
quark luminosity factor, ${\cal L}_q$, and the hard subprocess cross section,
${\hat \sigma}(\qq \to \gg)$
\be
\label{eq:q2}
 \sigma_q \; = \; {\cal L}_q (M_{\gg}^{2}, y) \: \hat{\sigma}_q (M_{\gg}^{2}).
\ee
We discuss the computation of these factors in turn.

\subsection{$\qq$ luminosity}

\begin{figure}
\begin{center}
\centerline{\epsfxsize=0.8\textwidth\epsfbox{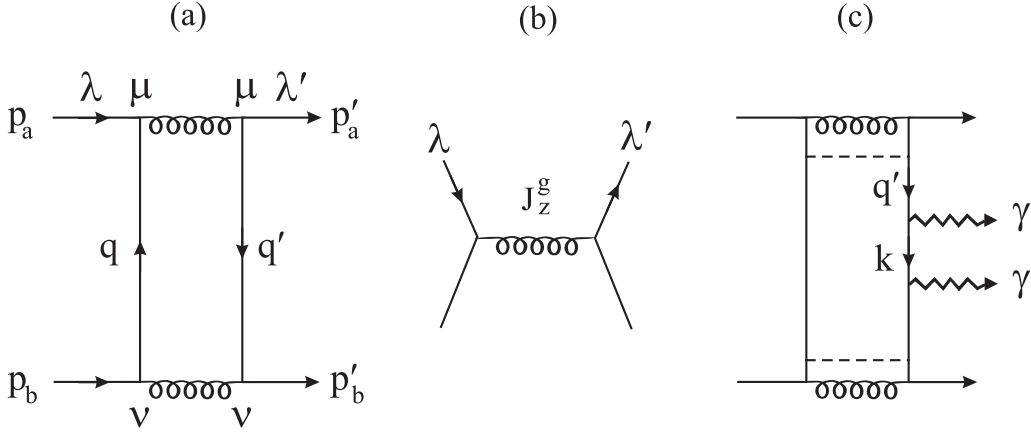}}
\caption{(a) The LO $\qq$ exchange diagram; (b) the helicity structure of the upper
part of diagram (a); (c) the inclusion of the hard subprocess.}
\label{fig:LO}
\end{center}
\end{figure}

To determine the luminosity, we first consider the leading-order $\qq$ exchange diagram,
Fig.~$\ref{fig:LO}$a, in the high energy limit.    Note that $s$-channel helicity
conservation, $\lambda=\lambda'$, holds for this process \cite{GGFL,LK}.  This may be seen from the Born
graph, Fig.~$\ref{fig:LO}$b, corresponding to the upper part of the diagram.  Due to helicity
conservation at each vertex, a fast incoming quark of $\lambda=+1/2$, say, produces a gluon with
$J_z^g=+1$, which then creates a quark with $\lambda'=+1/2$.  This property allows us to close the
external lines in Fig.~$\ref{fig:LO}$a, and to calculate the numerator of the amplitude as
\be
\label{eq:num}
{\rm Tr}[\slashp_a \gamma_{\mu} \slashq \gamma_{\nu} \slashp_b \gamma_{\nu} \slashq' \gamma_{\mu}]~=~
4{\rm Tr}[\slashp_a ~\slashq ~\slashp_b ~\slashq']~=~8sq_t^2.
\ee
Here we consider the forward amplitude with $p'_{at}=0$, and therefore $q_t=q'_t$.  Note that only
the transverse component, $q_t$, survives, since any longitudinal component of $q$ or $q'$
`annihilates' with $p_a$ or $p_b$ in (\ref{eq:num}), see \cite{GGFL,LK} for details.  That is the lowest-order luminosity
amplitude averaged over the incoming quark polarisations and colour indices is\footnote{Here
we consider the positive signature (singlet quark) exchange, where the real part of the
amplitude is small, $|{\rm Re} ~A_q/{\rm Im}~ A_q| \ll 1$.}
\be
\label{eq:B}
{\rm Im} ~A_q~=~\frac{16\pi^3}{N_C} \left(\frac{C_F}{2\pi} \alpha_s \right)^2 \frac{dq_t^2}{q_t^2},
\ee
where $C_F \alpha_s /2\pi$ represents the lowest-order unintegrated quark distribution,
given by the splitting function $P_{qq}(z)$ in the limit $z \to 0$.
After evolution of the parton densities, each factor $C_F \alpha_s /2\pi$ should be
replaced by the unintegrated distribution $f_q /x$.  Now consider the inclusion of the
hard subprocess, Fig.~$\ref{fig:LO}$c, in which two photons of mass $M_{\gamma \gamma}$
are produced.  The luminosity ${\cal L}_q$ in (\ref{eq:q2}), corresponding to $\qq$ exchange
with active quarks of a given flavour, is given by
\be
\frac{\partial{\cal L}_q}{\partial y \partial{\rm ln}M^2_{\gg}}~~=~~{\hat S}_q^2~\left(
\frac{2\pi}{N_C b}\int \frac{dq^2_t}{q^2_t M^2_{\gg}}
f_q(x_1,q_t^2,\mu^2)f_q(x_2,q_t^2,\mu^2) \right )^2.
\label{eq:A}
\ee
The unintegrated quark distributions,  $f_q$, are determined from
 the conventional quark densities by the relation
\be
f_q(x,q_t^2,\mu^2)~~=~~R_q ~\frac{\partial}{\partial{\rm ln}q_t^2} \left(xq(x,q_t^2)\sqrt{T_q(q_t,\mu)} \right),
\label{eq:fq}
\ee
in analogy with (\ref{eq:a6}).  Here the Sudakov factor is
\begin{equation}
\label{eq:Tq}
 T_q (q_t, \mu) \; = \; \exp \left ( - \int_{q_t^2}^{\mu^2} \:
 \frac{\alpha_S (k_t^2)}{2 \pi} \: \frac{dk_t^2}{k_t^2} \:
 \int_0^{1-\Delta}  P_{qq} (z) dz \right ),
\end{equation}
 which ensures no gluon emissions in the quark evolution from $q_t$
up to the hard scale $\mu$.
The $q_t^2$ in (\ref{eq:num})
is the reason why the $1/q_t^4$ in the analogous equation (\ref{eq:a4}) becomes $1/q_t^2$ in (\ref{eq:B})
and in (\ref{eq:A}).  The factor 2 in brackets reflects the fact that the hard subprocess may be initiated
by either the $t$-channel quark with momentum $q$ or $q'$.  The origin of the $1/M_{\gamma \gamma}^2$
arises from the $1/s$ suppression of the $\qq$-exchange amplitude in comparison with the two-gluon
exchange amplitude, together with the $1/x_1x_2$ factors from the $f_q/x$ noted above.

Strictly speaking the survival factor of the rapidity gaps, ${\hat S}_q^2$, in (\ref{eq:A})
 may be different from
the survival factor ${\hat S}_g^2$ in (\ref{eq:a4}) for the $gg$ exchange process, due to
the different impact parameter profiles of the quarks and gluons inside
 the incoming protons \cite{KMRsoft, KKMR}.
To account for the different profiles we may use different slopes $b$ in (\ref{eq:a4}) and (\ref{eq:A}).
However it turns out \cite{KMRS} that for realistic values of $b$ = 4--6 ${\rm GeV}^{-2}$, the product
${\hat S}^2/b^2$ is almost constant.   The value is ${\hat S}^2/b^2 = 3 \times 10^{-3}$ and $1.5 \times
10^{-3}~{\rm GeV}^4$ for the Tevatron and the LHC respectively.

Another difference may arise since each eigenstate of the multichannel eikonal model, with its own
absorptive cross section, may have its own parton
composition \cite{KKMR}. However we do not expect this difference to be significant for low $x$ partons.

\subsection{The subprocess cross section }

\begin{figure}
\begin{center}
\centerline{\epsfxsize=0.15\textwidth\epsfbox{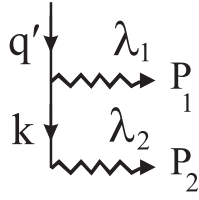}}
\caption{The extra subamplitude we have to calculate for 
the $q\bar q$ subprocess for $\gg$ production.}
\label{fig:subp}
\end{center}
\end{figure}

After the luminosity of (\ref{eq:A}) is calculated, the remainder of Fig.~$\ref{fig:LO}$c, corresponding to
 the amplitude of the hard subprocess, contains
the quark propagator $1/\slashk$, the vertices of the photon emissions,
and an additional quark propagator, say $1/\slashq'$, shown in the
schematic diagram of Fig.~$\ref{fig:subp}$.  The structure of the subamplitude is of the form
\be
\label{eq:amp}
\hat{{\cal M}}_{\lambda_1 \lambda_2}~~=~~\frac{\slashe_1~\slashk~\slashe_2}{k^2}~+~\frac{\slashe_2~\slashk_u~\slashe_1}{k_u^2},
\ee
where the additional quark propagator has been omitted for the moment.
The two terms correspond to the $t$- and $u$-channel
contributions.  The diagrams of Fig.~$\ref{fig:LO}$c and Fig.~$\ref{fig:subp}$ show the $t$-channel contribution, whereas the $u$-channel
amplitude corresponds to the permutation of the two photons.  Note that $k=q'-P_1$ and $k_u=q'-P_2$,
where $P_i$ are the momenta of the photons, $i=1,2$.  For the photon helicities,
we take the polarisation vectors
\be
\label{eq:pol}
-\vec {\epsilon}^{~\pm}_1,~\vec {\epsilon}^{~\mp}_2~~=~~i(\pm x' + iy')/\sqrt{2},
\ee
where the $x',y'$ plane is perpendicular to the photon momentum in the $\gamma\gamma$- rest frame.
In the massless quark limit, the amplitudes 
$\hat{{\cal M}}_{++}$ and $\hat{{\cal M}}_{--}$ vanish (see, for example, \cite{FKM}),
and so we need only consider $\hat{{\cal M}}_{\pm\mp}$.   Using $\slashe^+_1 \slashe^-_2=0$,
we find
\be
\label{eq:pm}
\hat{{\cal M}}_{+-}~~=~~\slashe_1~2(q'\cdot\epsilon_2)\left(\frac{1}{k^2}+\frac{1}{k_u^2}\right),
\ee
since $k=q'-P_1$ and $\epsilon_2\cdot P_1=0$ in the $\gamma\gamma$ rest frame.  Finally we have to convolute
$\slashe_1$ with the quark propagator $\slashq'_t/q'^2$, and so the spin structure of the subprocess amplitude
for the diagram of Fig.~$\ref{fig:subp}$ is given by
\be
\label{eq:subamp}
{\cal M}_{+-}~~=~~2(q'_t\cdot\epsilon_1)(q'\cdot\epsilon_2)
\left(\frac{1}{k^2}+\frac{1}{k_u^2}\right)\left(\frac{1}{q'^2}\right).
\ee

{}From the formal point of view, the result (\ref{eq:subamp}) may be regarded as the on-mass-shell
amplitude for $\qq \to \gg$ annihilation averaged over the quark colours and helicities,
together with a specific averaging over the transverse momentum, $q'_t$, of the incoming
quark, which we specify below. We now justify this statement.  First, at LO, $q'_t$ in (\ref{eq:A}) is much less than the
photon transverse momenta $P_t$.  In this limit the incoming quarks may be treated as
on-mass-shell fermions.  Second, the colour factor was already included in the colour singlet luminosity
(\ref{eq:A}), so we must average over the colours of the quarks.  Next, we have used unpolarised quark densities,
so we need to average over the helicities of the quarks.  The final `averaging' over $q'_t$ is more subtle.
Recall that in the calculation of the luminosity amplitude (\ref{eq:B}), only the transverse
momentum component $q_t$ survives in the $t$-channel quark propagator.
(This is not the case for the usual on-mass-shell $\qq \to \gg$ amplitude.)
As a consequence of $s$-channel helicity conservation for the incoming proton,
the projection of the total angular momentum of the produced $\gg$ system on the beam ($z$) axis
satisfies $J_z^{\gg}=0$.  Recall that we are considering forward proton scattering.  On the
other hand, due to quark helicity conservation, we have $J_z^{\qq}=\pm 1$.  So we require
quark orbital angular momentum to satisfy $L_z=\pm 1$, which reveals itself through $q'_t$.
In the limit $q'_t \to 0$, there is no way to generate $|L_z|=1$.
Angular momentum conservation kills everything which does not depend on $q'_t$.  Therefore the effective amplitude
should be written as a difference of two matrix elements which correspond to subprocesses with different quark beam
directions originating from their transverse momenta.  This leads to
\be
\label{eq:eff}
\frac 1{N_C} \sum_{i,k} \delta_{ik} ~\frac 12 \sum_{\lambda,\lambda'} \delta^{\lambda,-\lambda'}~
\frac{1}{2}\left[{\cal M}^{\lambda\lambda'}_{ik}(q'_t)~-~{\cal M}^{\lambda\lambda'}_{ik}(-q'_t)\right],
\ee
where $i,k$ and $\lambda,\lambda'$ are the quark colour and helicity indices respectively.  For simplicity,
elsewhere in the paper these indices have been omitted.
At first sight, this expression still appears to vanish once we integrate over the azimuthal
angle of $q'_t$.  Indeed this is true for a point-like amplitude $\cal{M}$; for example
for forward central exclusive production of a $Z$ boson.  However for our
non-local amplitude, there is a correlation between the direction of the quark $q'_t$
and the photon $P_t$.  Thus a contribution of $O(q'^2_t/P_t^2)$ survives after the
azimuthal angular averaging of $(q'_t\cdot\epsilon_1)(q'\cdot\epsilon_2)$ in (\ref{eq:subamp}).
This $q'^2_t$ cancels the factor $1/q'^2 \simeq 1/q'^2_t$ in (\ref{eq:subamp}), coming from the quark propagator,
and so finally we obtain the effective $\qq \to gg$ hard subprocess cross section\footnote{The behaviour
$d\hat{\sigma}/dt \propto (1/M_{\gg}^4)({\rm cos}\theta/{\rm sin}\theta)^4$
may be explained without
an explicit calculation, instead using arguments based on the rotation properties
of the amplitude and the Wigner $d$-functions. Indeed, for spinless
particles, $d\sigma/dt\propto r_T^4$, where the radius of interaction
$r_T\sim 1/P_{t\gamma} \sim 1/{\rm sin}\theta$. This is easy to
check in $\lambda\phi^3$ theory. Next we have to satisfy a set
of selection rules. We consider the photon helicity amplitude
with $(\lambda_1,\lambda_2)=(+,-)$ (or $(-,+)$), which has projection
$|J_{z'}|=2$  of the total $\gamma\gamma$ angular momentum
on the photon axis $z'$.  Simultaneously the projection of the total
$\gamma\gamma$ angular momentum on the quark axis, $z_q$, is
$J_{z_q}=\pm 1$. The probability amplitude for such a configuration
is given by ${\rm sin}\theta {\rm cos}\theta$.
  On the other hand the projection on the incoming
proton direction ($z$) is $J_z=0$. This can only be possible due to
the precession of the quark axis $z_q$ around the proton axis $z$
 with $|L_z|=1$. The probability amplitude to have $|L_z|=1$ is
proportional to $q_t r_T$, that is to the ratio of the quark and
photon transverse momenta $q_t/P_{t\gamma} \sim 1/{\rm sin}\theta$. Finally
we need to take the right sign of $L_z$,
 and to sum the contribution of the `$t$' and `$u$' channel diagrams.
Since the `$u$' channel is obtained by replacing $P_t$ by $-P_t$, this gives
another factor ${\rm cos}\theta$ (as in the difference of  the two
$d$-functions $d^1_{1,1}-d^1_{1,-1}={\rm cos}\theta$).
Thus we obtain an additional factor in the amplitude of
$({\rm cos}\theta/{\rm sin}\theta){\rm sin}\theta{\rm cos}\theta={\rm cos}^2 \theta$.
The behaviour $d\hat{\sigma}/dt \propto 1/M_{\gg}^4$ comes just
from dimensional counting. Therefore, including the first
kinematical factor of $1/{\rm sin}^4\theta$, we obtain $d\hat{\sigma}/dt\propto
(1/M_{\gg}^4)~({\rm cos}\theta/{\rm sin}\theta)^4$.}
\be
\frac{d\hat {\sigma}_{\rm eff}}{dt}~=~16\pi \left(\frac{e_q^2\alpha}{M^2_{\gg}} \right)^2
\left( \frac{{\rm cos}~\theta}{{\rm sin}~\theta} \right)^4.
\label{eq:sigqq}
\ee
Here $e_q$ is the electric charge of the quark and $\theta$ is the scattering angle in
the $\gg$ rest frame.

To calculate the observable cross section we have to include the contributions of the active
quarks and antiquarks of all flavours.   However the luminosity (\ref{eq:A}) is written for
the cross section for one type of quark.   To sum up all the quark contributions we must
sum the $\qq$ luminosity amplitudes ( given by the square root of the right hand side of (\ref{eq:A}))
multiplied by the amplitudes of the hard subprocess
 (and not the cross sections).  Finally we have accounted for the identity of the photons
 and summed over the (+,-) and (-,+) photon helicity configurations  in the cross
 section.

\section{Discussion of results}

\begin{figure}
\begin{center}
\centerline{\epsfxsize=\textwidth\epsfbox{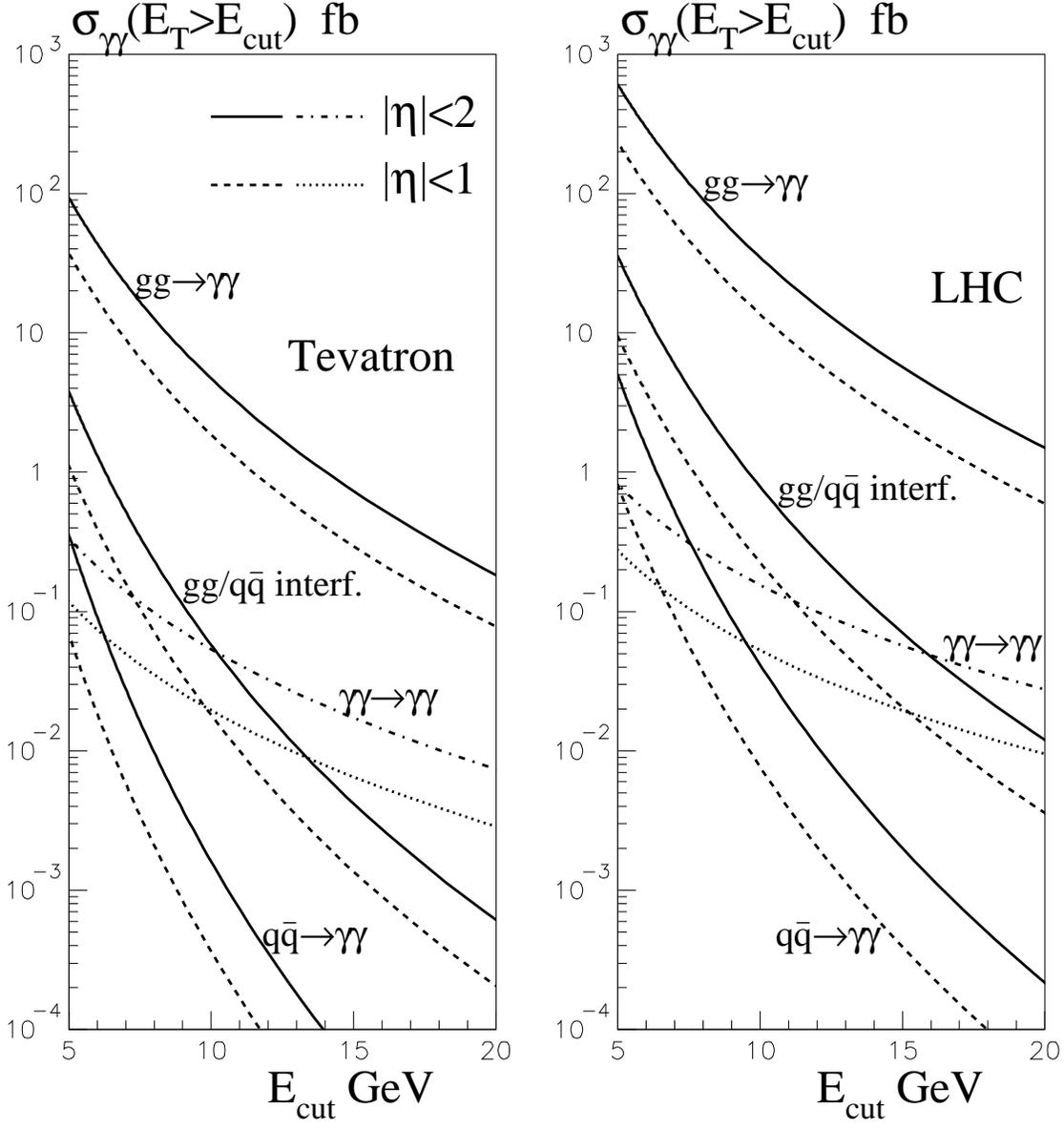}}
\caption{The contributions to the cross section for exclusive $\gg$ production
from $gg$ and $\qq$ exchange at the Tevatron and the LHC.   Also shown is the
contribution from the QED subprocess $\gg \to \gg$.  For each component we
show the cross section restricting the emitted photons to have $E_T>E_{\rm cut}$
and to lie in the centre-of-mass
rapidity interval $|\eta_{\gamma}|<1$ (or $|\eta_{\gamma}|<2$).}
\label{fig:results}
\end{center}
\end{figure}

Using the formalism described above, we have calculated the cross section of exclusive
$\gg$ production for the Tevatron ($\sqrt s =1.96~$TeV) and the LHC energy ($\sqrt s =14~$TeV).
In Fig.~$\ref{fig:results}$ we present the cross section integrated over the kinematic
domain in which the emitted photons have transverse energy $E_T>E_{\rm cut}$ and
centre-of-mass rapidity, for both photons, either $|\eta_{\gamma}|<1$ or $|\eta_{\gamma}|<2$.
Clearly the dominant contribution comes from $gg$ $t$-channel exchange.
In spite of the large enhancement coming from the skewed quark factor, $R_q^4 \sim 200$,
the contribution which originates from $\qq$ exchange is more than two orders of
magnitude lower; and falls more steeply with increasing $E_T$ due to the factor
$1/M_{\gg}^4$ in the luminosity.  Such a small $\qq$ exchange contribution to
exclusive $\gg$ production is explained, first, by the $(q'^2_t/M_{\gg}^2)$
suppression coming from angular momentum conservation, and, secondly, by the
${\rm cos}^4 \theta$ behaviour of the subprocess cross section.  The cross section
vanishes at 90 degrees, while the $\eta_{\gamma}$ cuts select events with small
 ${\rm cos} \theta$.
Calculating the interference between the $gg$ and $\qq$ exchange amplitudes,
 we account for the helicity structure of the hard subprocess amplitudes and
 for the complex phase of the $gg \to \gg$ amplitude.

 The results shown in Fig.~$\ref{fig:results}$ are obtained using MRST partons \cite{MRST99}.
The predictions differ by up to about 20\% if CTEQ partons \cite{CTEQ} are used; the
cross section being a little larger at the Tevatron and a little smaller at the LHC.
 
Recall that both the luminosities ${\cal L}_g$ and ${\cal L}_q$ were
 calculated for forward outgoing protons, that is in the limit of
 vanishing $p'_{at}$ and $p'_{bt}$.  This is a very good approximation for
 the $\qq$-exchange contribution, since the additional suppression factor
 $(q'^2_t/M_{\gg}^2)$, which is implicit in (\ref{eq:A}), in comparison
 with (\ref{eq:a4}), makes the $q_t$ integral logarithmic.  In addition, the
 Sudakov factor $T_q$ in (\ref{eq:fq}) pushes the dominant $q_t^2$ region,
 in the integral, closer to the factorization scale $\mu^2$.
 This justifies the use of the
 massless quark approximation to calculate the effective cross section (\ref{eq:sigqq}).
 Since the colour
 charge of the quark is smaller than that of the gluon, and the dominant
 $q_t^2$ interval is closer to $\mu^2$,
 the suppression of the $\qq$-exchange
 contribution of the cross section arising from $T_q$ ($\sim 0.6-0.8$)  is much weaker than
 the suppression of the $gg$-exchange component due to $T_g$.
 
 The corrections due to non-zero $p'_{it}$ of the outgoing protons, which
 are the order of $(p'_{it}/q_t)^2\sim 1/q_t^2b$, are quite small.  For the
 $gg$-exchange contribution, the saddle-point of the integral (\ref{eq:a4})
 is in the region $q^2_t \sim 1-1.5 ~{\rm GeV}^2$ for the Tevatron energy,
 and $q^2_t \sim 1.5-3 ~{\rm GeV}^2$ for the LHC energy, depending on the value
 of $E_T$.  Thus the violation of the $J_z=0$ selection rule may be as large as
 $(p'^2_{it}/q^2_t)^2\sim 10\%$.  

It is interesting to note that the fraction of $\qq$ induced events at the LHC is larger
than that at the Tevatron.  This is because the quark densities at relatively
large scales grow faster, with decreasing $x$, than the gluon densities at lower scales and low $x$.

\begin{figure}
\begin{center}
\centerline{\epsfxsize=0.7\textwidth\epsfbox{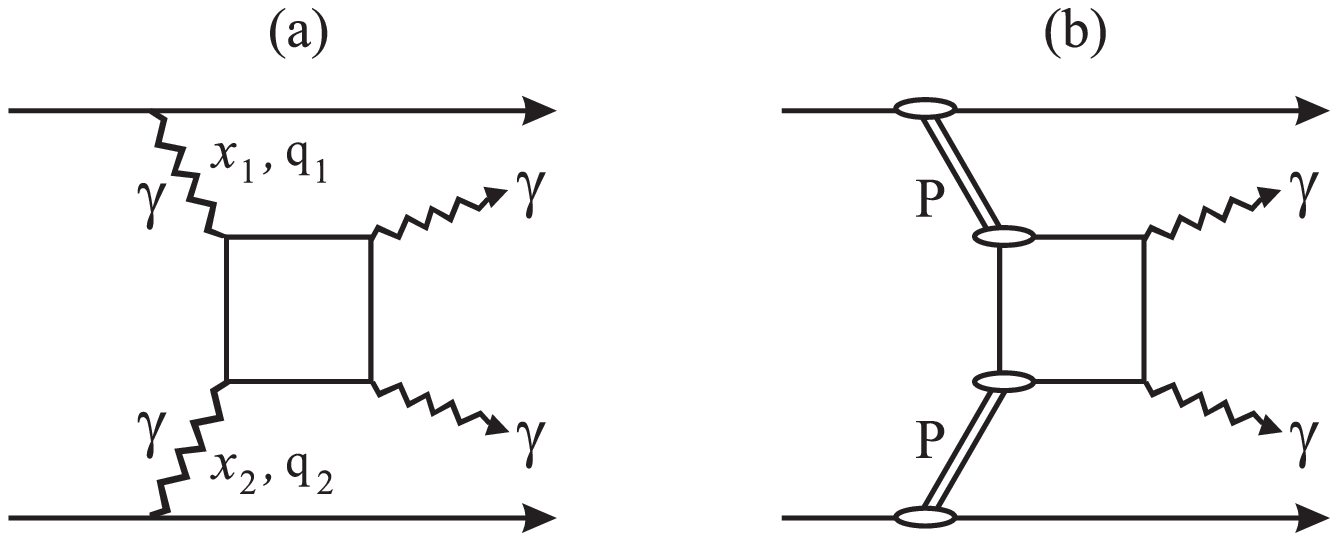}}
\caption{Large-distance contributions to exclusive $\gg$ production: (a) the QED
induced process, and, (b) the Pomeron-Pomeron fusion process.}
\label{fig:QED}
\end{center}
\end{figure}

To complete the discussion of the sources of exclusive $\gg$ events we consider
contributions originating from large-distance processes.  First we have the QED
process shown in Fig.~$\ref{fig:QED}$a.    The effective $\gg$ luminosity reads
\be
\frac{\partial{\cal L}_{\gg}}{\partial y \partial{\rm ln}M^2_{\gg}}~~=~~{\hat S}_{\gamma}^2~
\left({\frac{\alpha}{\pi}}\right)^2 \int_{q^2_{\rm min}}\frac{dq^2_1}{q^2_1} F^2_N(q^2_1)
\int_{q^2_{\rm min}}\frac{dq^2_2}{q^2_2} F^2_N(q^2_2),
\label{eq:QED}
\ee
with $q^2_{{\rm min},i}=x_i^2 m^2_p$, where $m_p$ is the mass of the proton.
The momentum fractions carried by the incoming photons
are
\be
x_{1,2}~=~(M_{\gg}/\sqrt s) e^{\pm y},
\ee
The $F_N$ are the usual dipole form factors of the proton.  They provide the upper cut-off
on the integrals.  To calculate the $\gg \to \gg$ amplitudes we use Ref.~\cite{[37], BFDCW}.  We include
fermion loops for the quarks, electron, muon and tau.  Since the $\gg$ luminosity
comes from large impact parameters, that is very low $q_i^2$, the survival factor
${\hat S}_{\gamma}^2 \sim 1$, see \cite{KMRphot}. The resulting exclusive QED
contribution to the $\gg$ cross section is shown in Fig.~$\ref{fig:results}$\footnote{
For large $E_T$, $E_T>110$ GeV, the QED contribution starts to dominate.  However the
cross section is very small, about $3 \times 10^{-4}$ fb for $|\eta_\gamma|<2$ at the LHC.}.

Next we consider the same process but with the $t$-channel photons replaced by
Pomerons, see Fig.~$\ref{fig:QED}$b.  We use this diagram to compute the low $q_t~(<0.85$ GeV) contribution to
the luminosity in (\ref{eq:a4}), which was excluded from the perturbative calculation.
We put the same limit on the virtuality of the $t$-channel left quark line
in Fig.~$\ref{fig:QED}$b.   This cut-off strongly reduces the size of the Pomeron-Pomeron$\to \gg$
amplitude.   Unlike exclusive $\chi_c$ production, where the perturbative and non-perturbative
contributions are comparable \cite{KMRS}, here the $\gg$ yield from Fig.~$\ref{fig:QED}$b is
less than a few percent of the perturbative $gg \to \gg$ cross section.  As noted in Ref.~\cite{KMRS},
there is no interference between the real amplitudes of the diagrams in Fig.~$\ref{fig:QED}$
and the imaginary amplitude corresponding to Fig.~\ref{fig:gg}a; here we refer to the luminosity
amplitudes, since for the $J_z=0$ case the fermion loop contribution is real.
Strictly speaking this is only true for the process Fig.~$\ref{fig:QED}$b if we assume that the
non-perturbative Pomeron interacts with the quark via a photon-
like vertex $\gamma_\mu$, which provides $s$-channel quark
helicity conservation. For the case of $\gamma$-exchange, Fig.~$\ref{fig:QED}$a,
there may be some $|J_z|=2$ contribution where the $\gamma\gamma
\to \gamma\gamma$ amplitude has its own imaginary part. However
this contribution, coming from large impact parameters $b_t$,
corresponds to  very low $p_{it}$ of the forward protons, and thus
essentially does not interfere with the main amplitude, Fig.~$\ref{fig:gg}$a.

\section{Conclusions}

The double diffractive exclusive production of a massive
system (such as a Higgs boson) is a good way to search, and to study, New Physics at
the LHC.   The existence of rapidity gaps on either side of the system means that the
event rate will be suppressed.   The observation of the exclusive production of a pair
of high $E_T$ photons at the Tevatron offers the possibility to check the exclusive prediction of these
types of process. As can be seen from Fig.~6, the dominant contribution to the
exclusive diffractive production of such a pair of photons is
driven by the same two-gluon exchange mechanism, that is by the same
effective $gg^{PP}$ luminosity, as is exclusive diffractive Higgs
boson production. Therefore, indeed, this process can be used as
a `standard candle' to check and to monitor the exclusive $gg^{PP}$ luminosity
that has been used for the prediction of the Higgs cross section.

The uncertainty of the predictions comes from the parton distributions used to calculate the
luminosities, the model dependent calculation of the survival factors ${\hat S}^2$ and the
lack of knowledge of the NLO corrections to the hard subprocess.  The first two have been discussed
above and in Ref.~\cite{KKMRCentr}.  Since we would like to use exclusive $\gg$ as a `standard candle'
to monitor the exclusive $gg^{PP}$ luminosity, it is important to calculate the NLO correction to $gg \to \gg$ amplitudes
accounting for the presence of the additional $t$-channel gluon (shown on the left in Fig.~$\ref{fig:gg}$a)
which provides the effective infrared cutoff for the NLO loop contribution.

%%%%%%%%%%%%%%%%%% perhaps a statement that we would believe the prediction for Higgs at the LHC
%%%%%%%%%%%%%%%%%% a bit more than \gamma\gamma at the Tevatron

\section*{Acknowledgements}

We thank Mike Albrow, Albert De Roeck, Risto Orava, and especially Beate
Heinemann and Angela Wyatt
for useful discussions and encouragement. 
ADM thanks the Leverhulme Trust for an Emeritus Fellowship and MGR thanks the IPPP at the University of
Durham for hospitality. This work was supported by
the UK Particle Physics and Astronomy Research Council, by a Royal Society special
project grant with the FSU, by grant RFBR 04-02-16073
and by the Federal Program of the Russian Ministry of Industry, Science and Technology
SS-1124.2003.2.

\end{document}